\documentclass[prb,aps,showpacs,preprintnumbers,twocolumn,superscriptaddress]{revtex4}
\input{epsf.sty}

\def\eq{\begin{equation}}
\def\eeq{\end{equation}}
\def\eqa{\begin{eqnarray}}
\def\eeqa{\end{eqnarray}}

\def\pref#1{(\ref{#1})}

\newcommand{\roughly}[1]{\mathrel{\raise.3ex\hbox{$#1$\kern-0.85em
\lower1ex\hbox{$\sim$}}}}

\begin{document}

\preprint{DIAS-STP-06-25}
\title{Modular Symmetry, the Semi-Circle Law and Quantum Hall Bilayers}

\author{C.P.~Burgess$^{1,2}$ and B.P.~Dolan$^{3,4}$\\
{\it $^1$ Department of Physics and Astronomy, McMaster University,\\
\qquad 1280 Main Street West, Hamilton, Ontario, Canada, L8S 4M1.}\\
{\it $^2$ Perimeter Institute, 31
Caroline Street North, Waterloo, Ontario, Canada, N2L 2Y5.}\\
{\it $^3$ Dept. of Mathematical Physics,
National University of Ireland, Maynooth, Ireland.}\\
{\it $^4$ School of Theoretical Physics, Dublin Institute for Advanced Studies,
10 Burlington Rd., Dublin, Ireland.}\\
{{\it e-mail:} cburgess@perimeterinstitute.ca,
bdolan@thphys.nuim.ie}}


\date{9th January 2007}

\begin{abstract}
There is considerable experimental evidence for the existence in
Quantum Hall systems of an approximate emergent discrete symmetry,
$\Gamma_0(2) \subset SL(2,Z)$. The evidence consists of the robustness
of the tests of a suite a predictions concerning the transitions
between the phases of the system as magnetic fields and
temperatures are varied, which follow from the existence of the
symmetry alone. These include the universality of and quantum
numbers of the fixed points which occur in these transitions;
selection rules governing which phases may be related by
transitions; and the semi-circular trajectories in the Ohmic-Hall
conductivity plane which are followed during the transitions. We
explore the implications of this symmetry for Quantum Hall systems
involving {\it two} charge-carrying fluids, and so obtain
predictions both for bilayer systems and for single-layer systems
for which the Landau levels have as spin degeneracy. We obtain
similarly striking predictions which include the novel new phases
which are seen in these systems, as well as a prediction for
semicircle trajectories which are traversed by specific
combinations of the bilayer conductivities as magnetic fields are
varied at low temperatures.
\end{abstract}


\maketitle
\section{Introduction}

Quantum Hall systems continue to provide remarkable laboratories
for studying the rich dynamics which is possible for
strongly-correlated electron systems in two spatial dimensions.
The now-classic studies of the many Hall plateaux found within
systems for which only the first Landau level is occupied
\cite{QHESLReview}, were later supplemented by the
discovery of new and remarkable phenomena in systems involving
higher Landau levels \cite{QHEHLLReview}, and the discovery of
evidence for new superfluid phases in bilayer systems, involving
two or more Quantum Hall layers which are spaced sufficiently
closely to allow inter-layer coherence and correlations
\cite{QHEBLReview}.

Although the microscopic picture of the electron physics in these
systems is well understood in terms of the Laughlin wave-function
\cite{LaughlinWF} --- at least for those systems for which only
the first Landau level is occupied --- it is also useful to have a
purely long-wavelength description of the effective degrees of
freedom which are relevant (in the renormalization-group -- RG --
sense) for the low-energy transport measurements. As is well-known
from the study of superconductors (for which the BCS theory can be
understood as the relevant effective description for energies near
the superconducting gap \cite{PS}), such a long-distance
description complements the microscopic picture by identifying the
domain of validity of the low-energy predictions and so thereby
allowing a better understanding of the robustness of various
low-energy phenomena.

For homogeneous Quantum Hall systems the large number of available ground
states (compared, for example to a BCS superconductor) complicates
the low-energy description of the theory, and the detailed form of
the effective theory applicable at mK temperatures is not yet
fully understood.  It has been suggested by a number of authors\cite{shapere,Andy}
that there is a large discrete group, $\Gamma_0(2) \subset SL(2,Z)$,
that maps between these ground states --- using complex conductivities, 
$\sigma = \sigma_{xy} + i \sigma_{xx}$, the map is
\begin{equation} 
\sigma\rightarrow \frac{a\sigma +b}{c\sigma +d}
\end{equation} 
where $a$, $b$, $c$ and $d$ are integers constrained by $ad-bc=1$ and $c$ is even.
This symmetry commutes with the scaling flow
in a sense that will be explained below.
Indeed, much of the data concerning the transitions between
plateaux follows very robustly simply from the existence of this
symmetry \cite{SemicircleUs}. (We call this a `symmetry' even
though it relates distinct phases of the system, rather than
commuting with the Hamiltonian describing low-energy fluctuations
within any one phase. We use this name because the group {\it is}
a symmetry of the scaling flow.) On the theoretical side this symmetry
was first proposed on both phenomenological \cite{Andy} and then
on more microscopic grounds \cite{KLZ}, before being derived as a
consequence of particle-vortex duality for the relevant charge
carriers \cite{PVD} and then on very general grounds for
two-dimensional conformal field theories \cite{Witten}. The
increasing generality of the assumptions which enter into these
derivations helps to explain the surprisingly broad domain of
validity of the observations, which apply
well beyond the immediate vicinity of the critical points to which
they were originally expected to be restricted. Indeed the more
recent derivations show these symmetries even go beyond the domain
of linear response \cite{NonlinearTh}, in agreement with
observations \cite{NonlinearExp}.

As yet, there has not been an exhaustive study of the implications of
this emergent discrete symmetry for the physics of 
spin-degenerate Landau levels \cite{BrianGamma2} or of
bilayer systems. It is the purpose of this paper to do so, and we
find that the symmetry has a number of observational implications,
including the following.

\medskip\noindent $\bullet$
{\it The Symmetry:} In the single-layer case there are two classes
of emergent symmetry, depending on whether or not the charge
carriers are related by symmetry transformations to bosons or to
fermions \cite{PVD}. The corresponding symmetry for the bilayer
system is determined by which of these symmetries applies to the
individual layers in the limit where they are independent. Given
that all known spin-split monolayer QHE systems appear to be described by the
fermionic symmetry, $\Gamma_0(2) \subset SL(2,Z)$, we concentrate for the most part in
this paper on the generalization of the purely fermionic symmetry.
By combining the symmetries of each separate layer with the
symmetry of layer interchange we are led to a particular discrete
subgroup $G \subset Sp(4,Z)$. A similar analysis of the group which is
appropriate to systems built from `bosonic' layers is
straightforward, and is also presented in the event that such
layers should be experimentally realized.

\medskip\noindent $\bullet$
{\it Low-Temperature Fixed Points:} Since the symmetry commutes
with the flow, it makes universal predictions for the kinds of
fixed points which are possible at low temperatures. 
Unlike the
monolayer case, there cannot be a one-to-one relation between the
experimental variables --- temperature, $T$,
magnetic field, $B$, inter-layer separation, $d$, and charge carrier density, $n_s$ --- and the
six conductivities which are possible for composite systems built
from bilayers (in principle there could be a one-to-one correspondence for
identical bilayers, with only four independent conductivities, but we are not aware
of any experimental data that systematically explores the full parameter range).
Not all phases can be explored
for a given bilayer sample as $T$, $B$ and $d$ are varied, and a
consequence of this observation is that the possible attractive
fixed points which are allowed by the symmetries can depend on
other microscopic factors, such as the relative importance of
tunnelling and inter-layer or intra-layer Coulomb energies. We
identify three classes of flow which are consistent with the
bilayer symmetry, and identify the universal low-energy fixed
points which each predicts. These include all of the known phases
which have been observed for these systems, with one describing
widely-separated mono-layers, and the other two corresponding to
correlated bilayers in two situations distinguished by the
relative size of the tunnelling and inter-layer Coulomb energies.

\medskip\noindent $\bullet$
{\it Semicircles:} In the single-layer case, when particle-hole symmetry
is present, knowledge of the
temperature flow implied the system moved on semi-circular
trajectories within the conductivity plane when magnetic fields
are varied at low temperatures. This conclusion has
generalizations to the three categories of flows described above.
In particular, for samples exhibiting particle-hole symmetry,
semicircular transitions are again predicted for
specific combinations of conductivities in the cases of
widely-separated layers and for bilayers with large tunnelling
energies. 

\medskip\noindent $\bullet$
{\it Selection Rules:} We show how the experimentally-successful
selection rule, $pq' - p'q = \pm 1$, for monolayer transitions
between plateaux characterized by Hall conductivities $p/q$ and
$p'/q'$ (in units of $e^2/h$), generalizes to bilayer systems. For
bilayers precisely the same selection rule holds (for particular
combinations of conductivities) for all three categories of flow
mentioned above.

\medskip\noindent $\bullet$
{\it Unresolved Layers:} It is often the case that the separate
conductivities are not measured independently for bilayer systems,
and we show in this case how the above predictions reduce to the
case where only the total conductivity is measured. This allows
the predictions of the symmetry to be applied more directly to
single-layer systems for which it is the electron spins (or other
labels) which distinguish the two \lq layers'. The predictions in this
case include the existence of even-denominator states and
the selection rule becomes $pq' - p'q = \pm s$,
where $s = 1$ or 2, depending on which of the three categories of
flow is under consideration.
For systems with particle-hole symmetry there should be
semicircular transitions between these states.


\section{Emergent Symmetry of the Low-Energy Theory}

In this section we briefly describe the properties of the emergent
symmetry in the bosonic and fermionic cases, starting first with a
restatement of what is known for a single conducting fluid (as
appropriate to a spin-split Landau level in a monolayer sample)
and then generalizing to the case of two fluids (as would apply to
a spin-degenerate layer or a bilayer consisting of two spin-split
Quantum Hall systems).

\subsection{Single-Layer Case}

It is the case of a spin-split single Landau level which has been
longest studied, and we follow here the description of
ref.~\cite{PVD}. In this reference it is argued that there are two
operations which commute with the scaling flow in the low-energy limit
of any system in two spatial dimensions for which the
quasi-particle charge carriers relevant to transport measurements
are weakly-coupled particles or vortices. These two
generators do not commute with one another and their repeated
application generates a large discrete group which resides within
$SL(2,Z)$.

In two dimensions particle statistics is described by an angle,
$\varphi$, which represents the phase, $e^{i\varphi}$, which the
particles acquire if they are adiabatically moved around one
another. The first of the symmetry operations which must commute
with the flow is simply the addition of $2\pi$ statistics flux
to the relevant charge-carrying quasi-particles, since this does
not change at all the accumulated phase.

The second symmetry generator arises when the quasi-particles
relevant for transport in some of the system's phases are
weakly-coupled vortices rather than charges. In this case the
similar kinematics of charges and vortices in 2 dimensions implies
the flow commutes with the operation of particle-vortex
interchange.

Ref.~\cite{PVD} shows that these two operations have a very simple
action on the electromagnetic response of the 2D system, which for
conductors and semi-conductors acts on the complex conductivities
$\sigma \equiv \sigma_{xy} + i \sigma_{xx}$ as a fractional linear
transformation:
\begin{equation}
    \sigma \to \frac{a \sigma + b}{c \sigma + d} \,,
\label{fraclin}
\end{equation}
where the integers $a$, $b$, $c$ and $d$ satisfy $ad-bc = 1$.
Because this is invariant under a change of sign for the constants
$a$ through $d$, this defines the group $PSL(2,Z)$. It is
convenient to express the transformations of this type in terms of
two standard ones:
\begin{equation}
    S(\sigma) = - \, \frac{1}{\sigma} \qquad \hbox{and} \qquad
    T(\sigma) = \sigma + 1 \,,
\end{equation}
which satisfy the identity $(ST)^3 = 1$. For instance, the
implications for the conductivities of the addition of $2\pi$
statistics flux to the charge carriers can be written in terms of
these generators as the combination $S T^2 S$, or $\sigma \to
\sigma'$, with
\begin{equation} \label{StatisticsFluxAddition}
    \frac{1}{\sigma'} =  \frac{1}{\sigma} - 2 \,.
\end{equation}
This transformation was first derived by Jain and collaborators
\cite{Jain}, although only in the context of $\sigma_{xy}$ and not on the
full uper-half complex plane.

The modular group has fixed points, in the sense that there are
a special set of points $\sigma_*$ for which there exists an
element $\gamma$ of the modular group that leaves $\sigma_*$ invariant,
$\gamma(\sigma_*)=\sigma_*$.  Quantum Hall plateaux are examples 
of such fixed points lying on the real axis, $\sigma_{*,xx}=0$,
with $\sigma_{*,xy}=p/q$ a rational number (the requirement that $q$ be odd
for semiconductors restricts the modular group to a subgroup as described
in section II.A.2 below).

It is the specific values taken by the integers $a$ through $d$
for the particle-vortex interchange which differs in the bosonic
and fermionic cases, so we consider each of these cases separately
after first discussing the implications of this symmetry for the scaling 
flow.

\subsubsection{Scaling Flow}

The requirement that the scaling flow of the quantum Hall effect
commute with this infinite
discrete group\cite{fixedpoints} strongly restricts the general form of the
flow in Quantum Hall systems.
The statement that the symmetry commutes with the flow means that
the scaling function, the logarithmic derivative of the conductivity 
along a flow line ${\cal F}(\sigma,\bar\sigma)=\frac{d\sigma}{ds}$
where $s$ is the logarithm of the scale,
is the same regardless of whether the modular group is applied before or after
taking the derivative, {\it i.e} 
\begin{equation}
{\cal F}(\sigma,\bar\sigma)=\frac{d\sigma}{ds}\quad\Leftrightarrow\quad
{\cal F}(\gamma(\sigma),\gamma(\bar\sigma))=\frac{d\gamma(\sigma)}{ds}.
\end{equation}
This proves to be a very strong restriction and
has the consequence that fixed points of the modular symmetry
must necessarily be fixed points of the scaling flow, {\it i.e.} if 
$\sigma_*$ is such that there
exists an element $\gamma\in\Gamma_0(2)$ which leaves $\sigma_*$ invariant,
$\gamma(\sigma_*)=\sigma_*$, then $\sigma_*$ is necessarily a fixed point
of the scaling flow and ${\cal F}(\sigma_*,\bar\sigma_*)=0$.\cite{fixedpoints}
				   
The original scaling arguments of Khmel'nitskii\cite{Scaling} and Pruisken\cite{Pruisken} 
were obtained 
from RG techniques, using an \lq effective' sample size L. 
In general the conductivity would be expected to be
a function of a number of parameters, such as the magnetic field, 
the temperature, the charge carrier density, the impurity density and the size.  
If the charge carrier and impurity densities
are fixed then we have $\sigma(B,T,L)$, but simple scaling says that $\sigma$ is
dimensionless (in units with $e^2/h=1$) so it only depends on two 
independent arguments
and can be written as $\sigma(B/L^\eta, T/L^\rho)$ for some pair of
exponents $\eta$ and $\rho$, or alternatively as 
$\sigma(B/T^{\eta'}, L/T^{\rho'})$ for a different pair $\eta'$ and $\rho'$.
One expects a different flow depending on whether the effective sample size $L$ is varied,
${\cal F}_L:=L{d \sigma\over dL}=-\eta B{{\partial\sigma\over\partial B}\vline}_T
-\rho T{{\partial\sigma\over\partial T}\vline}_B$, or the temperature $T$ is varied,
${\cal F}_T:=T{d\sigma\over dT}=
-\eta' B{{\partial\sigma\over\partial B}\vline}_L
-\rho' L{{\partial\sigma\over\partial L}\vline}_B$.
If $B$ is held fixed then
${\cal F}_L=-\rho T{\partial\sigma\over\partial T}$
and
${\cal F}_T=-\rho' L{\partial\sigma\over\partial L}$.

Anticipating the discussion of fermions and considering
figure 1 below,
where an example of the flow dictated by modular symmetry is shown,
we note that the topology of 
the flow does not require full knowledge
of the functional form of ${\cal F}$.  If figure 1 is taken to represent
${\cal F}_L$ one can multiply ${\cal F}_L$ 
by any function of $L$ which is non-vanishing away from the critical points
and the topology of the figure does not change.
Alternatively, if figure 1 represents ${\cal F}_T$
one can multiply ${\cal F}_T$ 
by any function of $T$ which is non-vanishing away from the critical points
and the topology of the figure does not change.
These two alternative views are quite compatible, indeed figure 1 could
represent both ${\cal F}_L$ and ${\cal F}_T$,
if their ratio is a non-vanishing real function of $T$ and $L$, 
and this is perhaps not unreasonable from a field theoretical 
point of view in $2+1$ dimensions. 
$L$ can be taken to be the size of the system, e.g. the
spatial extent of the system at low $T$, while in field theory $1/T$ would be
the temporal extent of the system, so the distinction between the two
functions is that of temporal versus spatial variations.
Indeed there is experimental evidence that $L\sim T^{-p/2}$
with $p=2$.\cite{DynamicalScaling}

The experiments of Murzin et al\cite{Murzin} are in remarkable agreement
with figure 1 below and
indicate that it is reasonable to consider the flows obtained from 
modular symmetry for semi-conductors as arising
from ${\cal F}_T$.
In the following we shall take the pragmatic attitude, given the experimental
data, that the flows represent ${\cal F}_T$.

\subsubsection{Fermionic Symmetry}

If the charge carriers for one of the system's phases are fermions
(or related to fermions by a symmetry transformation), then the
implications for the conductivities of interchanging a particle
with a vortex is $\sigma \to \sigma'$ with
\begin{equation}
    \sigma' = \frac{ \sigma - 1}{2 \sigma - 1} \,.
\end{equation}
The group generated by iterating this transition with that of
eq.~\pref{StatisticsFluxAddition} consists of all $PSL(2,Z)$
transformations for which $c$ is even, which defines the subgroup
$\Gamma_0(2)$. This group is generated by the two generators $T$
and $S T^2 S$.

When supplemented
with a symmetry interchanging particles and holes ({\it i.e.}
$\sigma \to 1 - \bar\sigma$) completely determines a particular class
of semicircular flow lines \cite{SemicircleUs}. Combining this
with a boundary condition at high temperatures (or vanishing
magnetic field, {\it etc.}) to fix the overall direction of flow
leads to the desired observational consequences.

\vtop{ \includegraphics{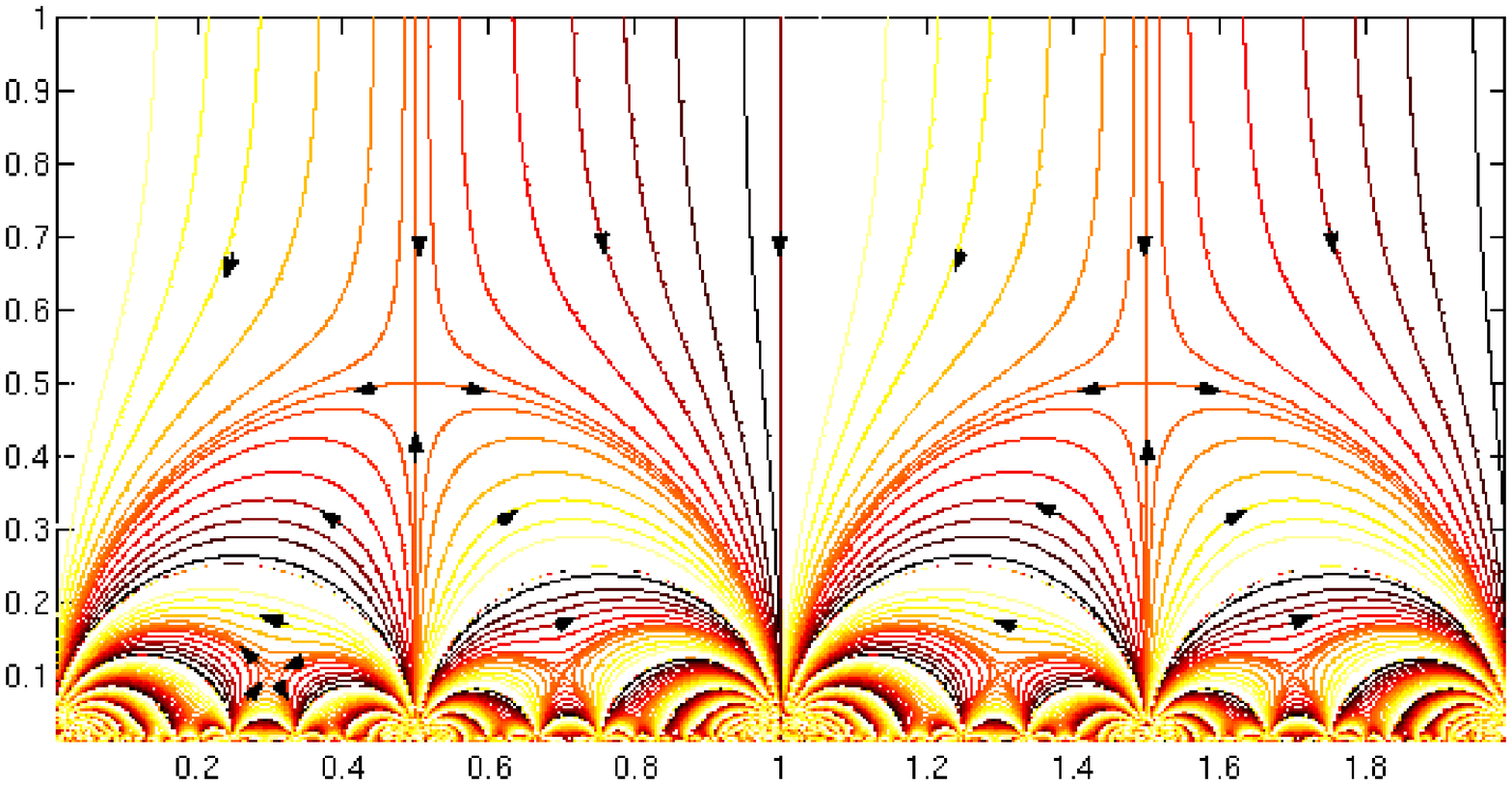} \vskip 5.8cm \centerline{\small Figure 1:
(Color online) Temperature flow lines for the $\Gamma_0(2)$-}  \vskip
0.005cm \centerline{\small invariant flow of $\sigma$. $\sigma_{xy}$ is plotted along the horizontal axis and} 
\centerline{\small $\sigma_{xx}$ up the vertical axis, in units with
$\scriptstyle e^2/h=1$. The arrows on the}
\centerline{\small figure indicate the flow direction for decreasing 
temperature,} 
\vskip 0.005cm \centerline{\small assuming the samples 
behave as monolayer semi-conductors} \vskip
0.005cm \leftline{\small when
$B=0$. $\phantom{XXx}$}}


\bigskip

For instance, assuming the flow direction at large $\sigma_{xx}$
(or at $\sigma_{xy} = 0$) corresponds to that of a semiconducting
monolayer sample leads to the temperature flow shown in Figure 1
(which is a more detailed version of figure 1 in the first reference of 
\ \cite{SemicircleUs}),
where the arrows indicate the direction of decreasing temperature
and different flow lines correspond to different values of $B$.
As the temperature is lowered the flow from above is forced 
onto semicircles of unit diameter spanning the integers on the real line
and, at low enough temperatures, these semcircles are followed when $B$ is varied
to force transitions between integer Hall plateaux.  This pattern is repeated
in a self-similar pattern for transitions between rational fractions.
 
There are completely attractive fixed points on the real line at rational fractions
$\sigma=p/q$ with $q$ odd (these are images of $\sigma=1$ under the group
action) and completely repulsive fixed points at even denominators, $\sigma=p/q$
with $q$ even (these are images of $\sigma=i\infty$).  There are saddle points,
attractive in one direction and repulsive in another, at $\sigma=(1+i)/2$ and its
images --- these represent critical points on the transition between two Hall
plateaux as the magnetic field is varied.

In the mathematical limit of perfect modular symmetry there is a fractal
structure just above the real line, but in any real sample there is obviously
a maximum value for the denominator of the filling fraction --- there can be no truly
infinite fractals in Nature --- though purer samples with
higher mobilities can achieve higher denominators.

Among the consequences of this flow are the following universal
properties
\begin{itemize}
\item[$M1$] The flow as temperatures go to zero is towards the
attractive fixed points, the complete list of which is $\sigma =
p/q$ with $q$ odd.
\item[$M2$] The plateaux, $\sigma = p/q$ and $\sigma' = p'/q'$,
which may be connected by a semicircle must satisfy the
selection rule $p \, q' - p'q = \pm 1$.
\item[$M3$] If particle-hole interchange
is also a symmetry, in the sense of creating
a hole when a particle is removed from a full 
Landau level ($\sigma \to 1-\overline\sigma$), then at low temperatures the flow between
plateaux obtained as magnetic fields are varied traverses a
semicircle in the complex conductivity plane, with the semicircle
centered on the real axis \cite{SemicircleUs}. This includes also vertical lines,
which may be regarded as infinitely large semicircles, leading to
the curves shown in Figure 1.
\end{itemize}

These describe very well the observed transitions which are seen
in the spin-split single-layer Quantum Hall systems in
semiconducting hetero-junctions.

\subsubsection{Bosonic Symmetry}

On the other hand, if the charge carriers for one of the system's
phases should be bosons (or related to them by a symmetry
transformation), then the group generated by iterating
Particle-Vortex interchange with $2\pi$ statistics flux addition
consists of all those $PSL(2,Z)$ transformations for which both
$b$ and $c$ are even, or for which both are odd. This defines the
subgroup $\Gamma_\theta$, which is generated by the two
generators $S$ and $S T^2 S$ (or equivalently $S$ and $T^2$). This
symmetry was first suggested in connection with the quantum Hall effect
in ref.~\cite{shapere}.

The requirement that the flow commute with this group
dictates its observational consequences
\cite{SemicircleUs}. These include the last two consequences
listed above for fermionic systems, but the first consequence is
instead changed to
\begin{itemize}
\item[$M1'$] The flow as temperatures go to zero is towards the
attractive fixed points, the complete list of which is $\sigma =
p/q$, where $p$ and $q$ are relatively prime and the product $p\,
q$ must be even.
\end{itemize}
Notice in this case that the selection rule M2 requires
transitions to be between plateaux for which $p/q =
\hbox{even}/\hbox{odd}$ and those for which $p/q =
\hbox{odd}/\hbox{even}$.

Because the fixed points differ between the cases of bosons and
fermions, the precise properties of the semicircles predicted in the bosonic and
fermionic cases differ in detail. Pictures of the allowed
semicircles in both cases are given in ref.~\cite{PVD}. Unlike for
the fermionic case, experimental systems exhibiting these
predictions were not known, although the recent observations of
Quantum Hall phenomena in graphene may provide the first example.\cite{graphene}

\subsection{Double-Layer Case}

For Quantum Hall bilayers there are more conductivities which are
accessible to observations, since there are Ohmic and Hall
conductivities for each layer separately, $\sigma_{11}$ and
$\sigma_{22}$, as well as the inter-layer conductivities,
$\sigma_{12} = \sigma_{21}$. Here the labels `1' and `2'
distinguish the two layers, and each of the $\sigma_{ij}$ are
complex numbers, whose real part gives the corresponding Hall
conductivity and whose imaginary part gives the Ohmic
conductivity.

Our interest is to generalize the discrete symmetry of the
single-layer case to the bilayer context, and it is convenient for
this purpose to group these complex conductivities into a complex
matrix
\begin{equation}
    \Sigma = \pmatrix{ \sigma_{11} & \sigma_{12} \cr
    \sigma_{21} & \sigma_{22} \cr} \,.
\label{Conductivity}
\end{equation}
In this case we know that the flow must commute with the action
of the discrete symmetry when it is applied to each layer
separately. If the layers are identical we may also demand that
the flow commute with the symmetry corresponding to the
interchange of the two layers. This leads us to expect there to be
a total of 5 independent symmetry generators in the bilayer case.

In terms of the matrix $\Sigma$ these transformations form a
subgroup of the discrete group
\begin{equation}
    \Sigma \to (A \Sigma + B) (C \Sigma + D)^{-1} \,,
\label{Sp4Z}
\end{equation}
where the $2\times 2$ matrices $A$, $B$, $C$ and $D$ have integer
entries. Denoting the transpose by the superscript `$t$', these
matrices must satisfy the conditions $A B^t = B A^t$, $C D^t = D
C^t$ and $A D^t - B C^t = 1$. The four matrices $A,B,C$ and $D$,
with this constraint, represent an element $\gamma=\pmatrix{A &
B\cr C & D \cr}$ of the group $Sp(4,Z)$.
A mathematical property of this transformation that is important
for the physics of the quantum Hall effect in bilayers is that,
when the imaginary part of $\Sigma$ is positive ({\it i.e.} its eigenvalues 
are positive) then (\ref{Sp4Z}) preserves this property.

Since the precise subgroup defined by the particle-vortex symmetry
for each layer depends on whether the pseudo-particle charge
carriers are related by the symmetry to bosons or fermions, the
same is also true for the bilayer case. We must ask for the
explicit form for the matrices $A$ through $D$ separately for the
case of bosonic and fermionic systems.

\subsubsection{Fermionic Symmetry}

For bilayers built from two fermionic layers the 5 symmetry
generators can be taken to be given by the following
transformations, which express the symmetry $\Gamma_0(2)$ acting
separately on each layer, together with interlayer interchange,
$P$. Collecting the matrices $A$ through $D$ into a $4 \times 4$
matrix
\begin{equation}
    \gamma \equiv \pmatrix{A & B \cr C & D \cr}
\end{equation}
the 5 generators are
\begin{eqnarray}
    &&T_1 = \pmatrix{ 1 & \frac12 (1 + \tau_3) \cr 0 & 1 \cr}
    \qquad
    S_1 T_1^2 S_1 = \pmatrix{ -\tau_3 & 0 \cr 1 + \tau_3 & -\tau_3 \cr}
    \nonumber\\
    &&T_2 = \pmatrix{ 1 & \frac12 (1 - \tau_3) \cr 0 & 1 \cr}
    \qquad
    S_2 T_2^2 S_2 = \pmatrix{ \tau_3 & 0 \cr 1 - \tau_3 & \tau_3
    \cr} \nonumber\\
    &&\hbox{and} \qquad P = \pmatrix{\tau_1 & 0 \cr 0 & \tau_1 \cr}\,.
\label{Bi-Fermions}
\end{eqnarray}
Here $\tau_k$ denotes the usual $2\times2$ Pauli matrices, with
\begin{equation}
    \tau_1 = \pmatrix{0 & 1 \cr 1 & 0 \cr} \qquad
    \hbox{and} \qquad
    \tau_3 = \pmatrix{1 & 0 \cr 0 & -1} \,.
\end{equation}
Most of the remainder of this paper is devoted to the exploration of the
observational consequences of this symmetry.

\subsubsection{Bosonic Symmetry}

Before exploring the consequences the fermionic bilayer symmetry,
we pause briefly to record the symmetry which would apply to
bilayers built from two bosonic layers. In this case the 5
symmetry generators can instead be found by replacing the
generators $T_1$ and $T_2$ by the following two generators
\begin{eqnarray}
    &&S_1 = \pmatrix{ \frac12 (1 - \tau_3) & \frac12 (1 + \tau_3) \cr
    -\, \frac12 (1 + \tau_3) & \frac12 (1 - \tau_3) \cr}
    \nonumber\\
    &&S_2 = \pmatrix{ \frac12 (1 + \tau_3) & \frac12 (1 - \tau_3) \cr
    -\, \frac12 (1 - \tau_3) & \frac12 (1 + \tau_3) \cr}
    \,.
\label{Bi-Bosons}
\end{eqnarray}
For simplicity we here do not further explore the consequences of
this group.

\subsection{Identical Layers}

In order to explore the consequences of the fermionic bilayer
symmetry, $G$, it is useful to specialize to the experimentally
relevant case where both of the layers are identical. In this case
we may write the intra-layer conductivities as $\sigma_{11} =
\sigma_{22} \equiv \sigma$, and the inter-layer conductivities as
$\sigma_{12} = \sigma_{21} \equiv \tilde\sigma$, and so
\begin{equation}
    \Sigma = \pmatrix{ \sigma & \tilde\sigma \cr
    \tilde\sigma & \sigma \cr} = \sigma \, \tau_0
    + \tilde \sigma \, \tau_1 \,,
\label{ConductivityIdent}
\end{equation}
where $\tau_0 = I$ denotes the $2\times 2$ unit matrix. With this
choice the generator $P$ acts trivially on $\Sigma$.

In this case the implications of the remaining four generators of
$G$ are more conveniently identified by following its action on
the following combinations $\sigma_\pm := \sigma \pm
\tilde\sigma$, in terms of which we have
\begin{equation}
    \Sigma = \sigma_+ \, \tau_+ + \sigma_- \, \tau_- \,,
\end{equation}
where $\tau_\pm$ denote the projection matrices $\tau_\pm =
\frac12(\tau_0 \pm \tau_1)$. Physically, the conductivities
$\sigma_\pm$ describe correlations of the currents $J_\pm = J_1
\pm J_2$, where $J_i$ denotes the electrical current passing
through layer `$i$'.

The action of $G$ on these variables is most simply expressed by
also focussing on the matrices $A$ through $D$ which can also be
written in terms of $\tau_\pm$: $A = a_+ \, \tau_+ + a_- \,
\tau_-$, and so on. The integers $a_\pm$, $b_\pm$, $c_\pm$ and
$d_\pm$ obtained in this way must satisfy $a_n d_n - b_n c_n = 1$
separately for $n = +$ and $n = -$. Similarly, the matrix $C$
obtained in this way is even provided the constants $c_\pm$ are
also even.

Using these matrices it is clear that the action of $G$ on
$\Sigma$ decomposes into a separate $\Gamma_0(2) \subset SL(2,Z)$
action on each of $\sigma_\pm$ --- denoted $\Gamma_0^\pm(2)$ ---
with
\begin{equation}
    \sigma_n \to \frac{ a_n \, \sigma_n + b_n}{c_n \, \sigma_n +
    d_n} \,, \qquad n = \pm \,.
\end{equation}
It is also clear that these two copies of $\Gamma_0(2)$ commute with
one another.

The advantage of writing things this way is that it allows the
results of ref.~\cite{PVD} to be taken over in whole cloth, since
this action is mathematically identical to the action of the
fermionic symmetry on a monolayer. In particular, we see that
writing the symmetry in this way implies that as the temperature
is varied, the topology of the flow is given as in
ref.~\cite{PVD}. There are then potentially two qualitatively different flow
patterns, which differ only in the direction of the flow along the
lines drawn in Figure 1.

When the flow with decreasing temperature is in the same direction
as is shown in Figure 1 then an initially large Ohmic conductivity
falls with temperature, as is the case for the semiconducting
monolayer systems. Continuity implies that this direction of flow
also holds in the limit of bilayers which consist of two
monolayers which are sufficiently far apart from one another. This
need no longer be true if a transition should occur to a
qualitatively different ground state at a critical separation,
such as happens once inter-layer correlations become important.

The curves followed when magnetic field, $B$, (or carrier density,
$n$) are varied at low temperatures can also be inferred from
temperature-flow diagrams like Figure 1. For these purposes notice
that different vertical lines correspond to different initial
magnetic fields, and (given the direction of flow given in Figure
1) if these start from the top of the figure (as is true for
comparatively low-mobility samples) then they are attracted to
integer plateaux (such as those at $\sigma = 0$ and $\sigma = 1$
in Figure 1) as $T \to 0$. By comparing the position for differing
magnetic fields at a common low temperature once can see that the
trajectories followed as $B$ is varied for fixed $T$ near zero lie
along the separatrix defined by the semi-circle connecting
$\sigma = 0$ to $\sigma = 1$ in Figure 1.\cite{PVD} In this case
the existence of the symmetry implies the following
generalizations of the monolayer predictions to bilayer systems.
\begin{itemize}
\item[$B1$] The flow as temperatures go to zero is towards the
attractive fixed points, the complete list of which is $\sigma_\pm
= p_\pm/q_\pm$ with $q_\pm$ odd.
\item[$B2$] The plateaux, $\sigma_\pm = p_\pm/q_\pm$ and $\sigma'
= p'_\pm/q'_\pm$, which may be related by such a semicircle must
satisfy the selection rule $p_n q'_n - p'_n q_n = \pm 1$
separately for $n = +$ and $n = -$.
\item[$B3$] At low temperatures (if particle-hole interchange is a
symmetry) the flow between plateaux obtained as the magnetic field
are varied traverses a semicircle in the complex $\sigma_+$ and a semi-circle
the complex $\sigma_-$ plane, with the semicircles centered on the real axis in each case.
This last prediction assumes the two plateaux in question lie in
the same in super-universality class --- an exception to this semicircle rule arises if the
varying magnetic field triggers a change in the category of the 
super-universality class of the transitions (as
we discuss in more detail below).
\end{itemize}

On the other hand, bilayers states have been seen which 
exhibit properties more usually associated with superconductors
rather than semiconductors and for these states an initially
large conductivity would be expected to get larger as the temperatures drops, rather
than lower as in a semiconductor.
For these states the flow for decreasing temperature is again along the lines of Figure
1, but with the direction of the flow lines reversed from those
drawn in the figure. The existence of such a flow is consistent
with the symmetry, and cannot be excluded on continuity grounds if
the bilayer system in question is separated from the limit of
well-separated layers by a discontinuous transition.

Indeed, we shall argue below that real bilayer systems do exhibit
low-temperature fixed points of this new form, including examples
\cite{Suen,Eisenstein} for which $\sigma_+ = \frac14$ and others \cite{Kellogg}
for which the resistivity matrix is given by
\begin{equation}
    -\Sigma^{-1} = \pmatrix{1 & 1\cr 1 & 1\cr} \,.
\end{equation}
These states, which arise for small inter-layer tunnelling
energies, are believed to involve a superfluid condensation of the
current corresponding to opposite-charge flow in the two layers,
and so is necessarily separated from the limit of widely-separated
layers by a phase transition. We return to a more detailed
discussion of the phenomenology of these new flows below.

For any such a phase having reversed flow direction, repeating the
arguments of ref.~\cite{PVD} leads instead to the following
predictions
\begin{itemize}
\item[$B1'$] The flow as temperatures go to zero is towards the
attractive fixed points, the complete list of which is $\sigma_\pm
= p_\pm/q_\pm$ with $p_\pm$ odd and $q_\pm$ even.
\item[$B2'$] The plateaux, $\sigma_\pm = p_\pm/q_\pm$ and $\sigma'
= p'_\pm/q'_\pm$, which may be related by such a trajectory must
satisfy the selection rule $p_n q'_n - p'_n q_n = \pm 2$
separately for $n = +$ and $n = -$.
\item[$B3'$] At low temperatures, if particle-hole interchange is 
a symmetry,  the flow between plateaux
obtained as the magnetic field is varied traverses a semicircle in
the complex $\sigma_+$ and a semi-circle in the complex $\sigma_-$ plane.
\end{itemize}
As discussed above, both of the latter two predictions require
there to be no discontinuous transition as $B$ is varied which
could change the category of super-universality class of symmetries 
that is relevant.

We see in this way the possibility of two different
categories of transitions amongst bilayer systems (and two more
if $\Gamma_\theta$ should also be a relevant group), with each
being governed by a qualitatively different kind of flow. In both
cases the trajectories of the conductivities has the shape given
by Fig.~1, but the two categories differ in the direction of this
flow as the temperature falls. The above predictions apply to the
transitions between phases, provided that the phases involved both
belong to the same category.

\subsection{Two Layers Unresolved}

More often than not experiments do not separately distinguish the
currents and voltages in any one layer, and quote only a partial
measurement corresponding to the total current and voltage across
both layers. This is the case both for {\it bona fide} bilayer
systems as well as for spin-degenerate monolayers for which it is
electron spin which plays the role of layer number. Because these
types of measurements are so common, we pause here to state what
the above symmetry statements become in this important special
case.

If the layers (or spins) are not distinguished observationally,
then all that is measured is the total conductivity, $\sigma_T$,
which is related to the inter- and intra-layer conductivities by
$\sigma_T = \sigma_{11} + \sigma_{22} + \sigma_{12} +
\sigma_{21}$. In the case of identical layers this simplifies to
\begin{equation}
    \sigma_T = 2(\sigma + \tilde\sigma) = 2 \,\sigma_+ \,.
\end{equation}

Since we have seen that the bilayer symmetry generates a
$\Gamma_0(2)$ action of $\sigma_+$ onto itself, we see that it
also must generate an action on $\sigma_T$. Interestingly, because
of the factor of 2 in the relationship between $\sigma_T$ and
$\sigma_+$, the action generated on $\sigma_T$ is {\it not} simply
$\Gamma_0(2)$. Instead, it is the transformation $\sigma_T \to (a
\sigma_T + b)/(c \sigma_T + d)$ with $b$ even, which defines the
group $\Gamma^0(2) \subset SL(2,Z)$.\footnote{The notation is that of
N.~Koblitz, {\it Introduction to Elliptic Curves and
Modular Forms.  2nd Edition}, Graduate Texts in Mathematics, No.97, Springer-Verlag (1984)
} This can be seen because $\Gamma^0(2)$ is
generated by $T^2$ and $STS$ and $\sigma_T\rightarrow \sigma_T+2$
and $\sigma_T\rightarrow {\sigma_T/( 1-\sigma_T)}$ are clearly
equivalent to $\sigma_+\rightarrow \sigma_++1$ and
$\sigma_+\rightarrow {\sigma_+/( 1-2\sigma_+)}$ when
$\sigma_T=2\,\sigma_+$.

We see that the total conductivity, $\sigma_T$, of unresolved
bilayer systems enjoys a $\Gamma^0(2)$ symmetry, which has
slightly different consequences than does $\Gamma_0(2)$. There are
again two cases, depending on the overall direction of the flow as
the temperature decreases. In particular, if the flow lines are
the same as for a semiconducting monolayer, predictions $B1-B3$
for $\sigma_+$ imply the following for $\sigma_T$:
\begin{itemize}
\item[$S1$] The flow as temperatures go to zero is towards the
attractive fixed points, the complete list of which is $\sigma_T =
p/q$ with $p$ even and $q$ odd.
\item[$S2$] The plateaux, $\sigma_T = p/q$ and $\sigma' = p'/q'$,
which may be related by varying magnetic fields at low
temperatures must satisfy the selection rule $p \,q' - p' q = \pm
2$.
\item[$S3$] For samples exhibiting particle-hole symmetry
the flow between plateaux obtained by varying the magnetic field 
at low temperature 
traverses a semicircle in the complex $\sigma_T$ plane, with the
semicircle centered on the real axis.
\end{itemize}
As usual the latter two of these predictions requires there is no
discontinuous change of ground state as the transition is made. In
particular, continuity implies that these properties hold in the
limit of widely-separated monolayers, for which interlayer
coherence and correlations are negligible.

On the other hand, for transitions within the category for which
the direction of the flow lines is reversed compared with Fig.~1
we have instead the predictions $B1'-B3'$ for $\sigma_+$, which in
the present context become:
\begin{itemize}
\item[$S1'$] The flow as temperatures go to zero is towards the
attractive fixed points, the complete list of which is $\sigma_T =
p/q$ with $p$ odd.
\item[$S2'$] The plateaux, $\sigma_T = p/q$ and $\sigma'_T =
p'/q'$, which may be related at low temperatures by varying the
magnetic field must satisfy the selection rule $p\, q' - p' q =
\pm 1$.\footnote{Recall here that $\sigma_+ =$ odd/even so we may
write $\sigma_+ = k/2\,l$ where $k$ is odd. Then $\sigma_T = 2\,
\sigma_+ = k/l$, and so $k'(2\,l) - k(2\,l') = \pm 2$ implies $k'l
- k \, l' = \pm 1$.}
\item[$S3'$] For systems with particle-hole symmetry 
the trajectory traversed when varying $B$ at low
temperatures follows a semicircle centered on the real axis.
\end{itemize}

\subsection{Transitions With Varying $B$}

We see that, when particle-hole symmetry is present, 
the condition that the flow commute with the
bilayer symmetry predicts a generalized semicircle law
for the trajectories which the conductivities
take at low temperatures as $B$ is varied. However, since changes
in $B$ can also drive transitions between the different categories
of phases of the bilayer systems, it is important to keep in mind
that the semicircular trajectories need not follow if the system
undergoes a transition from one category of phases to another. In
order to understand the implications of the above predictions for
real systems we must therefore keep track of how the phase of the
system changes as the magnetic field, $B$, (or magnetic length,
$\ell \propto B^{-1/2}$) varies.

\subsubsection{Relevant Energy Scales}

Physically, we expect that the kinds of phases which are possible
for a given system depends on the relative strength of a typical
inter-layer tunnelling matrix energy, $\Delta$, compared with the
inter- and intra-plane Coulomb correlation energies: $V_c^{12}$
and $V_c^{11}$. In terms of these quantities we expect the
following three regimes, which are schematically pictured as a
`phase diagram' in Figure 2 and discussed experimentally 
in\cite{Shayeganetal1},\cite{Shayeganetal2}:

\begin{itemize}
\item When $\Delta$ dominates both $V_c^{11}$ and $V_c^{12}$ then
inter-layer tunnelling is large and the number of electrons in
each layer is not separately conserved. In this case the Fermi
level can fall between the energies of the states whose
wave-functions are symmetric and antisymmetric under inter-layer
interchange, leading to what is effectively a single-layer system,
denoted ``Single Monolayer'' in figure 2.
If $\Gamma_0(2)$ is the relevant group then the older predictions,
$M1-M3$ or $M1'-M3'$, given above which apply.  This would be expected to be the case
at small $B$.
\item If $V_c^{11}$ is sufficiently dominant over $V_c^{12}$ and
$\Delta$ then we expect the system to behave as two independent
layers, denoted \lq\lq Independent Layers'' in the figure. In this case
we expect the direction of the temperature flow to be the same as
for each monolayers separately, and if $\Gamma_0(2)$ is the
relevant group then it is the predictions $B1-B3$, above, which
apply. (If the conductivities in each layer are not distinguished,
then it is instead the predictions $S1-S3$ which would be
relevant.)
\item When $V_c^{12}$ and $V_c^{11}$ are both more important than
$\Delta$, then inter-layer Coulomb correlations play a role and we
expect new bilayer plateaux to appear, denoted \lq\lq Correlated Layers''
in figure 2. This includes the situation
of Bose-Einstein condensation among the exciton states which form
from pairs of electrons and holes in different layers. In this
situation the existence of the superfluid state leads us to expect
the temperature flow to be opposite to that of the decoupled
monolayers, because conductivity decreases in a semi-conductor as
the temperature is lowered and increases in a superconductor.
This leads to the expectation that predictions $S1'-S3'$
are the relevant ones.
\end{itemize}

\vskip 0.4cm
\vtop{ \includegraphics{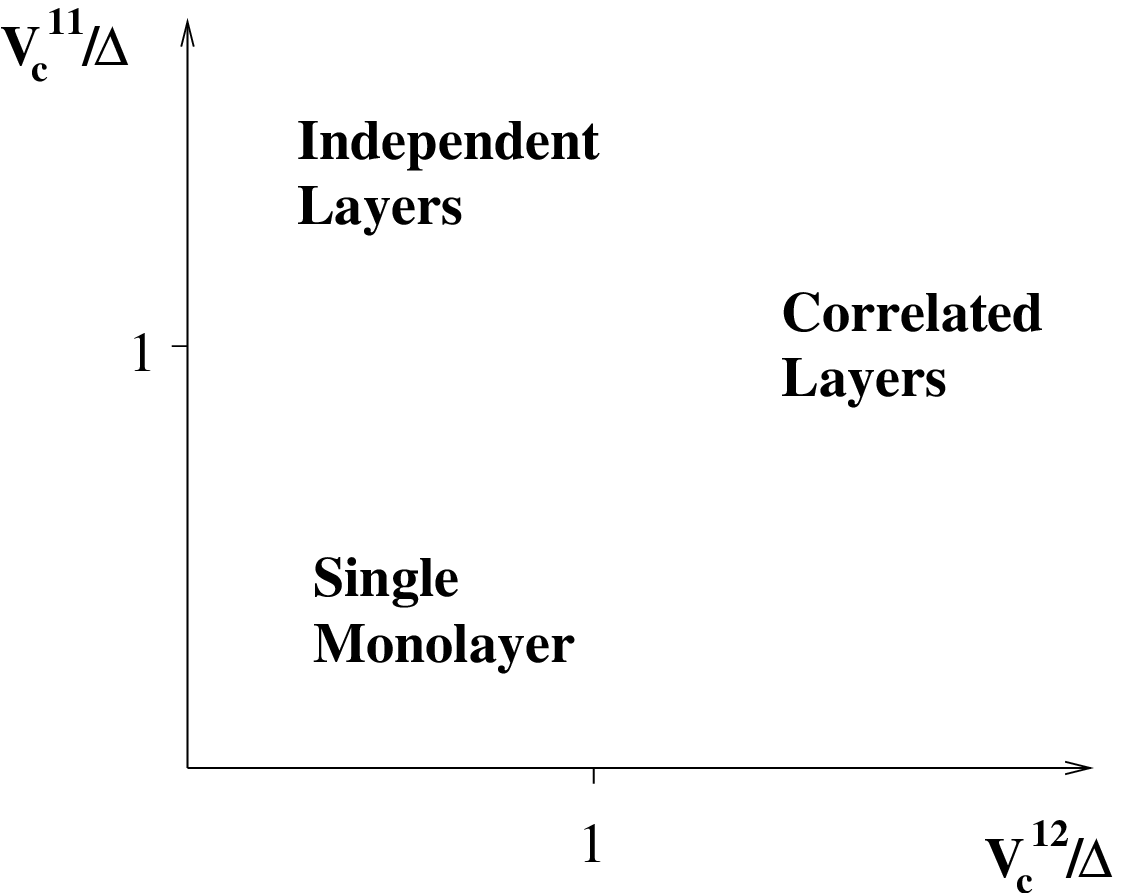} \vskip 6.2cm \centerline{\small
Figure 2: A sketch of the `phase diagram' describing the} \vskip
0.005cm \centerline{\small three regimes described in the text as
functions of intra-} \vskip 0.005cm \centerline{\small layer and
inter-layer Coulomb correlation energies norm-} \vskip 0.005cm
\centerline{\small alized to the tunnelling energy, $\Delta$.
$\phantom{xxxxxxxxxxxxxxxxx}$} }
\vskip 0.4cm
\bigskip

So far this entire discussion applies just as well to
spin-degenerate single-layer QHE systems as it does to spin-split
bilayers, provided that we identify $\Delta$ as the
Zeeman-splitting matrix element which is responsible for the
non-Coulomb part of the energy difference between the differing
spin states.

\subsubsection{Magnetic Field Dependence}

Because the energies $\Delta$, $V_c^{11}$ and $V_c^{12}$ all
depend differently on the magnetic field strength, any particular
sample traverses a trajectory in the phase diagram of Figure 2 as
$B$ is varied. It is here that we must distinguish between bilayer
and spin-degenerate monolayer systems, because they differ in the
$B$-dependence which is likely for these energy scales. This
difference has important implications for the kinds of
trajectories which are possible in Figure 2 for each type of
system as the magnetic field is varied.


\medskip\noindent{\it Bilayer Systems:} \medskip

\noindent For bilayers $\Delta$ is not expected to depend strongly
on $B$, but it does depend strongly on the inter-layer distance
$d$, falling to zero exponentially quickly as the layers are
separated.  The way in which the phase structure can depend on
magnetic field was examined in \cite{MacDonald} and figure 3 is
a sketch of figure 1 from that reference.  That analysis indicates
that, at large $B$, 
the ratio of the two length scales $d$ and the magnetic length 
$l\propto 1/\sqrt{B}$ determines whether or not the system behaves
as two independent monolayers or as a correlated bi-layer, as shown in 
figure 3 below.  As the magnetic 
field increases, at constant charge carrier density, 
the importance of the intra-layer
Coulomb energy ($V^{11}_c\propto 1/ l \propto \sqrt{B}$) 
relative to $\Delta$ increases until intra-layer Coulomb interactions
destroy the
correlated state and the system behaves as independent bi-layers,
provided $V_c^{11} >> V_c^{12}\sim 1/d$.

\vtop{ \includegraphics{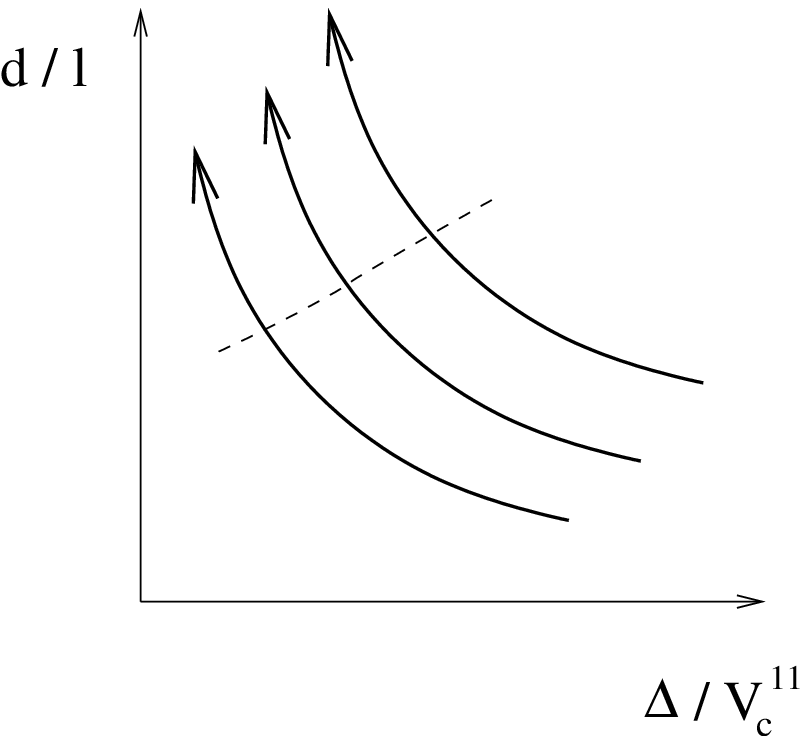} \vskip 7.3cm \centerline{\small Figure 3:
An illustration of the relative strengths of the intra} 
\centerline{\small -layer Coulomb 
energy and $\Delta$ compared to the dimensionless} 
\centerline{\small ratio $\scriptstyle d/l$, taken
from \cite{MacDonald}.  The curves correspond to increasing}  
\centerline{\small $\scriptstyle B$ at constant charge carrier density 
and the correlated bi-}  
\leftline{\small layer state is
only realised below the dotted line.}}
\bigskip

We see from these considerations that for semiconducting bilayers
at small magnetic fields $\Delta$ should always dominate over Coulomb
energies, implying
a description in terms of a single monolayer, for which the
predictions $M1-M3$ would be expected to hold. As $B$ increases
the Coulomb correlation energies become more important and, if
$V_c^{12}$ becomes important enough, a transition to the correlated
bilayer regime can occur (for which predictions like $B1'-B3'$ can
apply), depending on the detailed properties of the sample in
question. Eventually for large enough $B$ we expect $V_c^{11}$ to
dominate, leading to the behaviour of independent bilayers, for
which we expect $B1-B3$ to be relevant.

As we shall see, this sequence of pictures seems to provide a good
description of the kinds of phases which are seen in real Quantum
Hall bilayers, although more detailed measurements to confirm or
refute this picture would be most welcome.

\vskip 20pt
\noindent{\it Spin-Degenerate Monolayers:} \medskip

\noindent A similar discussion can be made for spin systems, with
two provisos: ($i$) that the conductivities of the two spins of
electron are not separately measured; and ($ii$) that some of the
relevant energy scales depend differently on $B$ than they do for
bilayer systems.

For spin systems the quantities $V_c^{12}$ and $V_c^{11}$ measure
the spin-dependent matrix elements of the Coulomb interaction in a
single plane, and so for the purposes of the qualitative estimates
in this section we take both to be proportional to $1/\ell \propto
B^{1/2}$. For spin systems we also take the matrix element
$\Delta$ to be due to the Zeeman interaction, and so $\Delta
\propto B$.

\vtop{ \includegraphics{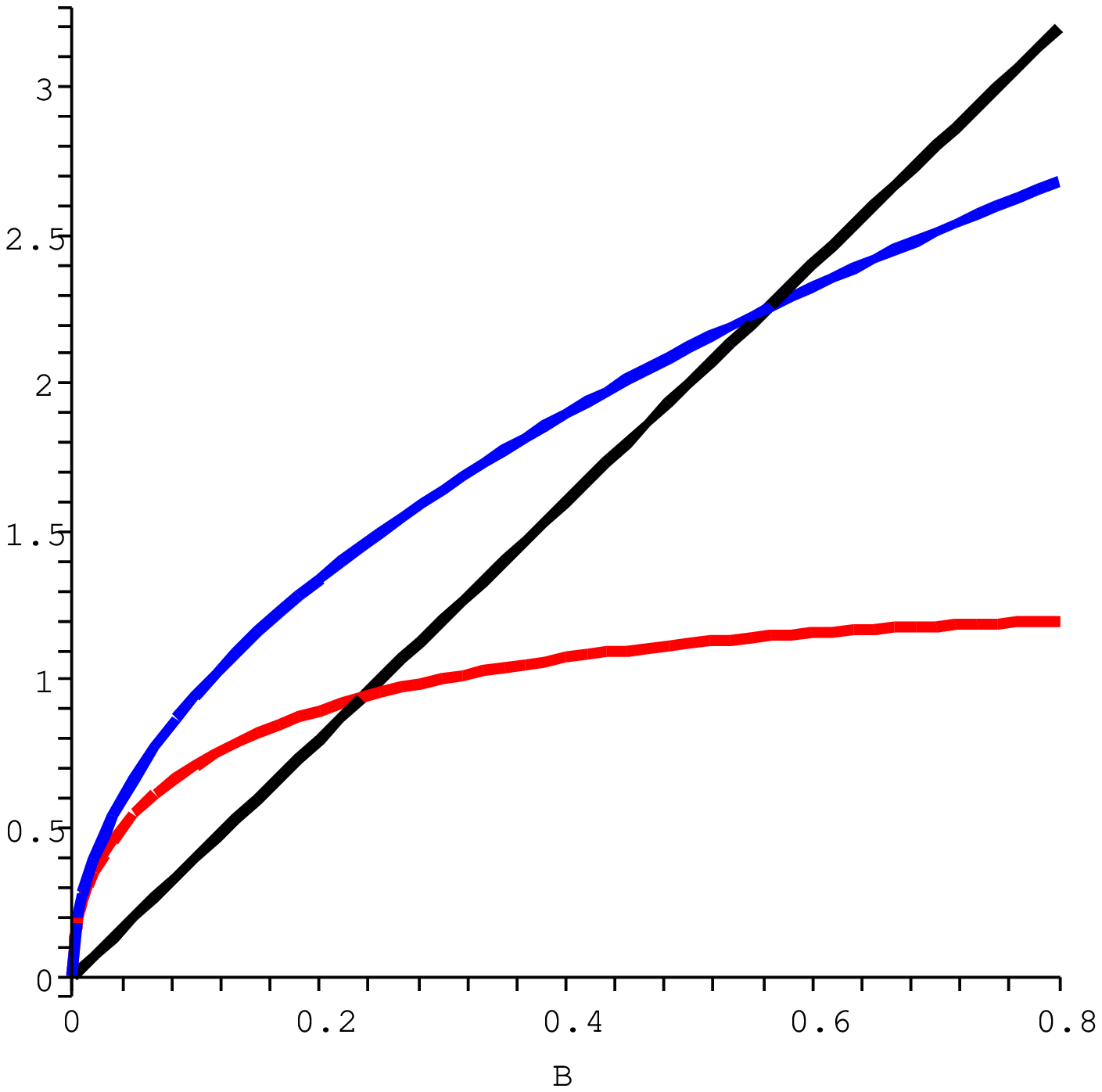} \vskip 9.3cm \centerline{\small Figure 4: 
(Color online) A sketch of the energies $V_c^{11}$} \vskip
0.005cm \centerline{\small (upper) and $V_c^{12}$ 
(lower) for monolayer systems with} 
\vskip 0.005cm \centerline{\small  spin-degenerate electrons.
The straight line shows}
\vskip 0.005cm \centerline{\small the Zeeman splitting, $\Delta$.} }

\smallskip 
Figure 4 gives a sketch of the relative size of these energies as
functions of $B$. In the case sketched the small-$B$ regime is
dominated by the Coulomb energies, assuming a sample with high
enough mobility that Quantum Hall plateaux are visible at all.
Assuming $\Gamma_0(2)$ is the relevant group, this leads to the
predictions $S1-S3$ (or $S1'-S3'$), depending on the relative
importance of $V_c^{11}$ and $V_c^{12}$.\footnote{Alternatively,
if $\Gamma_\theta$ is the relevant group -- as appears to be the
case for graphene -- the analogs of $S1-S3$ are required for this
group \cite{graphene}.} For larger $B$ the energy $\Delta$
eventually dominates to the extent that our simple $B$
parameterization continues to hold, leading to a monolayer
behaviour, with predictions $M1-M3$.

\medskip

\section{Applications}

We now explore how the above observations can be used to make
nontrivial predictions for real Quantum Hall systems. What makes
this tricky is the fact that when $B$ is varied, real systems 
can cross over between
the different kinds of categories of phases (as
described above) for which different symmetry predictions apply.
Such transitions are problematic because we only know that the
predictions hold for transitions amongst plateaux within any one
of these categories, and not for transitions in between different
categories.

In order to see how the above arguments may be used in practice,
we now examine their implications in more detail for several bilayer
systems for which experimental results are available in the
literature. Our strategy for doing so relies on the fact that for
each category described above there are three different nontrivial
predictions: ($i$) those governing the kinds of plateaux allowed
({\it e.g.} $B1$, $S1$, $B1'$ or $S1'$); ($ii$) those governing
the selection rules which express the plateaux that may be
obtained from any given one by varying $B$ ({\it e.g.} $B2$, $S2$,
$B2'$ or $S2'$); and ($iii$) those describing the shape of the
trajectories which the conductivities take as $B$ is varied
between these plateaux ({\it e.g.} $B3$, $S3$, $B3'$ and $S3'$).
Our strategy is to use the kinds of plateaux to identify the
category of phases which are present for each value of $B$. Once
these categories are identified in this way we may then test the
existence of the symmetry by asking whether the other two
predictions hold for transitions within each category.

\subsection{Unresolved Bilayers}

Most of the bilayer systems which have been studied experimentally
fall into our `single layer' category, for which only the total
conductivity, $\sigma_T$, is measured. For the purposes of this
section we take the example of refs.~\cite{Suen,Eisenstein}, which
together provided the first evidence for new kinds of Hall
plateaux in these systems having $\sigma_T = \frac12$. Both of
these references provide explicit traces of the total Ohmic and
Hall resistances at low temperatures as functions of magnetic
field. A representative trace as given in ref.~\cite{Suen} is
reproduced here as Figure 5.

At weak fields a series of integer plateaux are
seen with $\sigma_T = 5,4,3$ culminating in the first fractional
state at $\sigma_T = \frac83$. This is what is expected when
tunnelling dominates, including the appearance of the fractional
state once the Coulomb interaction becomes large enough to compete
with the others. Throughout this region we expect predictions
$M1-M3$ to apply, including the selection rule $|pq'-p'q| = 1$
(which can be seen from Fig.~5). The success of the picture 
can be tested on samples with particle-hole symmetry due to
the prediction $M3$ of a semicircle law for these transitions,
for which measurements would be most welcome.

Although all of the plateaux visible in Fig.~5 have odd
denominators, they cannot all be explained by the regime for which
$\Delta$ dominates. Instead a transition occurs to the
independent-bilayer category, for which predictions $S1-S3$ apply.
The plateaux in Fig. 5 with $\nu>8/3$ do
indeed have even numerators, but this in itself is
not sufficient to distinguish between 
$M1$ and $S1$.
That the
transition to $S1-S2$ occurs is instead seen in Fig.~5 from the change to the
selection rule $|pq'-p'q|=2$ for the plateaux above $\sigma_T =
\frac83$, indicating that $S2$ applies rather than $M2$. 
For particle-hole symmetric samples $S3$
the prediction of a semicircle law again applies, but one which links
plateaux obeying the new selection rule ($S2$) rather than the old
one ($M2$).


\vtop{\includegraphics{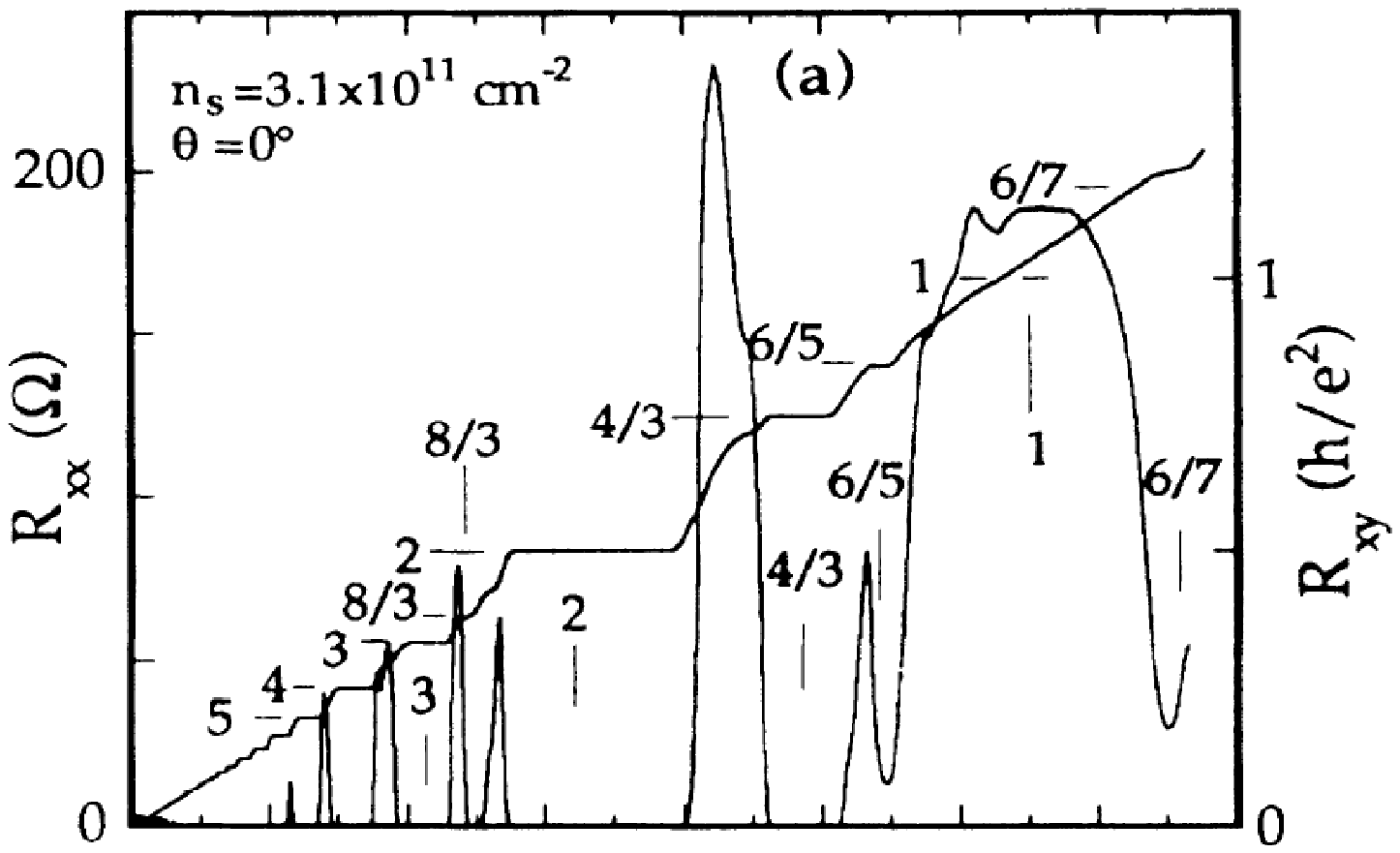} \vskip 5.9cm \centerline{\small
Figure 5: Resistance traces for a bilayer system, reproduced}
\vskip 0.005cm \centerline{\small \hskip -17pt  from Fig. 2 of Suen {\it
et.al.} \cite{Suen}. $B$ increases to the right in} \vskip 0.005cm
\centerline{\small \hskip -5pt this plot, with the same scale as for Fig.~7.
$\phantom{xxxxxxxxxxx}$}}
\bigskip

Of all the visible plateaux the selection rule $|pq'-p'q| = 2$
only fails for the transition from $\sigma_T = \frac65$ to
$\sigma_T = \frac67$, but there also appears to be considerable
structure between these two plateaux (including the incipient
state at $\sigma_T = 1$ --- more about which below) which
indicates unresolved intervening physics.

Figure 6 displays the plateaux found for the same sample but with
a smaller carrier density, $n_s$ (obtained by applying a different
bias voltage). As discussed in ref.~\cite{Suen}, this lower
carrier density implies a lower effective inter-layer separation,
$d$, leading to nontrivial interlayer Coulomb correlations.
For such a regime the predictions $S1'-S3'$ would apply, implying
that the observed plateaux should have $\sigma_T = p/q$ with $p$
odd, as well as the re-emergence of the selection rule $|pq'-p'q|
= 1$.
This picture appears to be borne out by Fig.~6, which differs from
Fig.~5 mainly in the appearance of the new state at $\sigma_T =
1$. Such a state agrees with the prediction $S1'$ but would be
forbidden for the independent-bilayer states which surround it,
since these must all have $\sigma_T = p/q$ with $p$
even.\footnote{More evidence that the $\sigma_T = 1$ plateau is
due to inter-layer correlations comes from its fragility to
tilting the applied magnetic field \cite{GMReview}.} And it is
further supported by the re-appearance of the selection rule
$|p q'-p'q| = 1$ for the transitions $\frac65 \to 1$ and $1 \to
\frac45$ (or the $\frac67$ state indicated in figure 6,
though there is no well developed plateau for that state), 
as dictated by prediction $S2'$. A gold-plated signal
would be the observation of the trajectory taken in the complex $\sigma_T$
plane during these transitions, since this should be described by
the semicircle trajectories predicted by $S3'$ in samples exhibiting particle-hole symmetry.

\vtop{\includegraphics{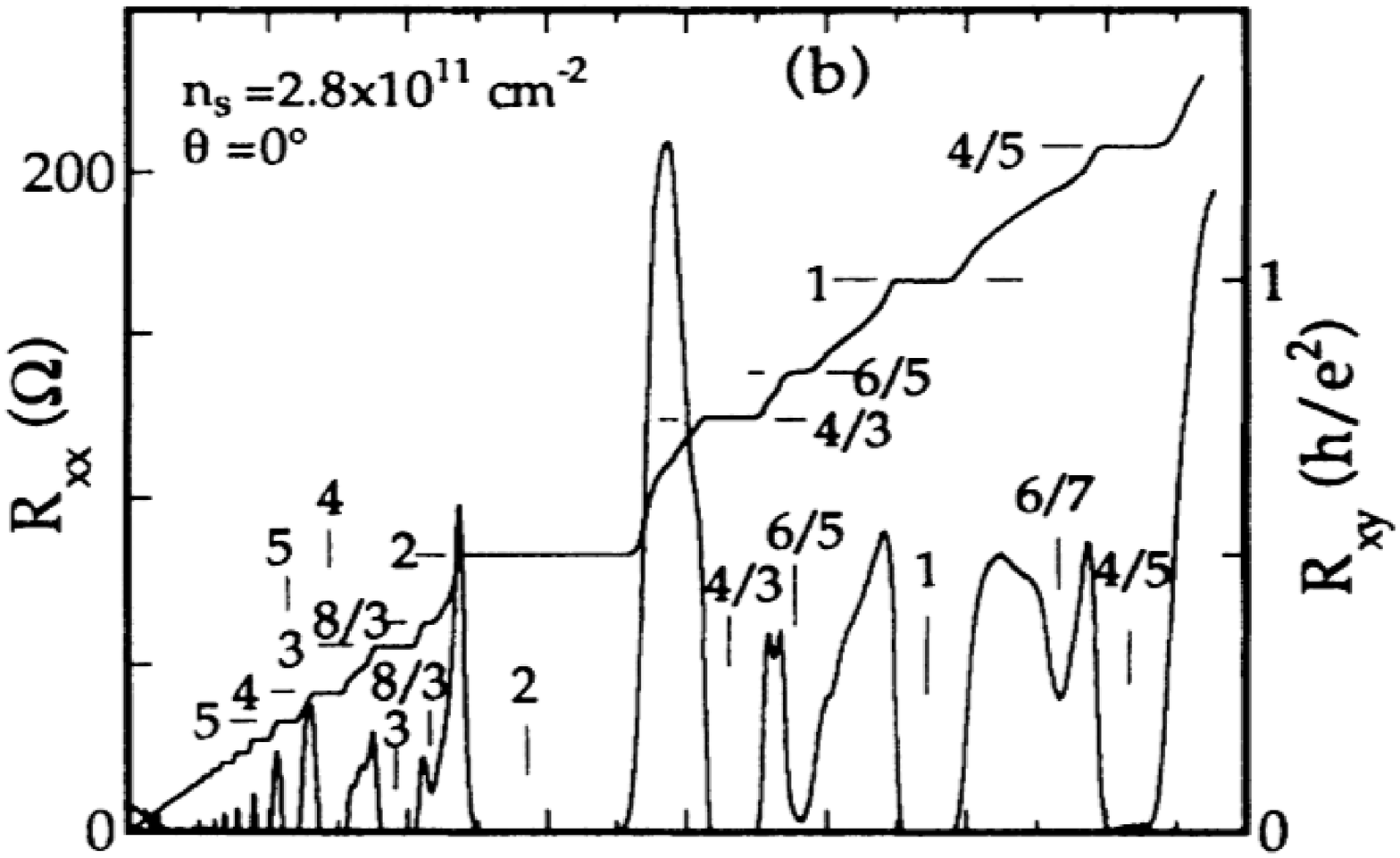} 
\vskip 6.2cm \centerline{\small
Figure 6: Resistance traces, from Fig. 2 of Suen {\it et.al.}
\cite{Suen},} \vskip 0.005cm \centerline{\small \hskip -10pt for the same
bilayer system as Fig.~5 but with smaller} \vskip 0.005cm
\centerline{\small  \hskip -15pt carrier density, $n_s$, and so smaller effective
interlayer} \vskip 0.005cm \centerline{\small separation,
$d$.}}
\bigskip


\vtop{ 
\includegraphics{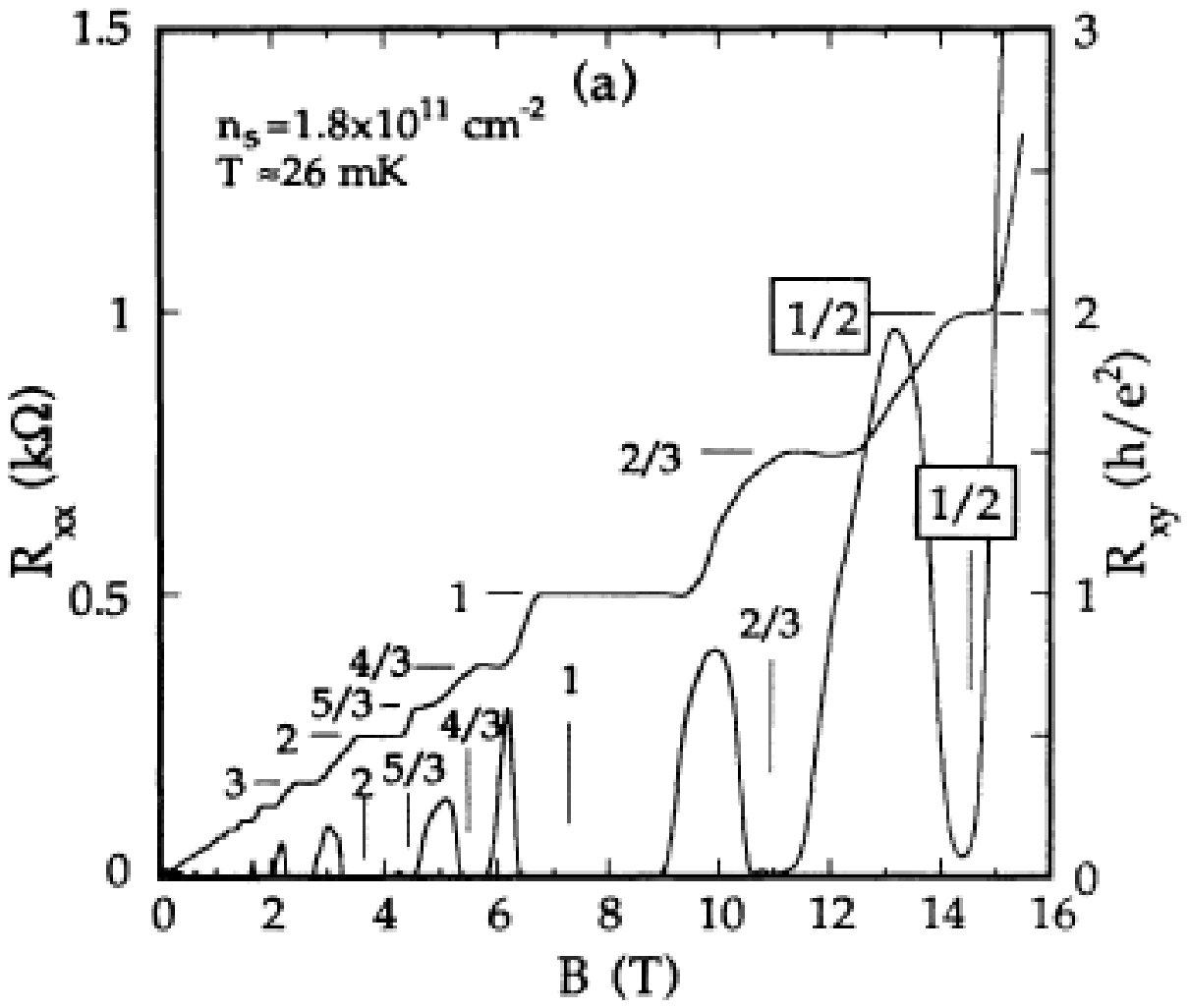} 
\vskip 7.2cm \centerline{\small
Figure 7: Resistance traces for the same bilayer system as} \vskip
0.005cm \leftline{\small  in Figs.~5 and 6, but with still
smaller $n_s$ and so smaller eff-} \vskip 0.005cm
\leftline{\small ective $d$. Data taken from Fig. 3 of Suen {\it
et.al.} \cite{Suen}. $\phantom{xxxxx}$}}
\bigskip

For still smaller $d$ the behaviour of the sample continues to
evolve. This may be seen from the resistance traces given in
Figure 7, which differ considerably from those of Figs. 5 and 6.
This figure agrees in many ways with what would be expected from
the trajectory of Fig.~3, including a phase of interlayer Coulomb
correlations, although there are also a few troubling aspects to
this description, as we now describe.

As discussed earlier, we qualitatively expect that shrinking $d$
causes $\Delta$ to grow, and this should increase the value of the
field, $B_c$, at which the transition occurs from
$\Delta$-domination to domination by $V_c^{11}$. We also expect
the regime of independent bilayers to become narrower, and perhaps
disappear, depending on how quickly $\Delta$ grows compared with
$V_c^{12}$.

Whether these are borne out by Fig.~7 depends on the regime with
which each of the plateaux is associated. Indeed all of the
observed plateaux in Fig.~7, except for the one at $\sigma_T =
\frac12$, occur at fractions with odd denominators and so could be
due to a tunnelling-dominated monolayer ($M1-M3$). On the other
hand, both plateaux $\sigma_T = \frac43$ and $\frac23$ have even
numerators and so could instead be interpreted as being due to the
independent-bilayer category of phases ($S1-S3$). These two
options can be distinguished using the selection rule ($M2$ or
$S2$), and since Fig.~7 indicates this to be $|p\,q'-p'q| = 1$ ---
apart from the $\frac53 \to \frac43$ transition, see below --- it
is the monolayer option which appears to be most appropriate.
Under this interpretation there would be a direct transition
between $\sigma_T = \frac23$ and $\sigma_T = \frac12$ from the
monolayer regime ($M1-M3$) to the regime of inter-layer Coulomb
correlations ($S1'-S3'$). If this picture is right then
monolayer-type semicircle trajectories ($M3$) are predicted for
all of the transitions in particle-hole symmetric samples, 
at least up to $\sigma_T =1$.

The $\frac53 \to \frac43$ transition is the only exception to the
selection rule $|p\,q'-p'q| = 1$, and the interpretation of this
transition is problematic within the present framework. However
this could indicate the existence of a poorly-resolved state in
between these two plateaux, similar to the way the incipient
$\sigma_T = 1$ state causes a funny selection rule between the
$\frac65$ and $\frac67$ states in Fig.~5. A more baroque
interpretation takes the $\frac53 \to \frac43$ transition to mark
the point where the transition is made into the
independent-bilayer regime ($S1-S3$), in which case Fig.~7 would
require a further transition to the correlated bilayer regime
($S1'-S3'$) for the $\sigma_T = 1$ state, followed by a re-entrant
independent-bilayer state at $\sigma_T = \frac23$. Such a
picture would however also require a second transition to the correlated
regime to account for the $\sigma_T = \frac12$ state and we shall not
pursue this possibility further here.

This analysis shows how the fact that there are three kinds of
predictions may be used both to reconstruct the regime which is
relevant, and to test the implications of the underlying
$\Gamma_0(2)$ symmetry of each Quantum Hall layer.

\subsection{Resolved Bilayer Systems}

The previous discussion involved only measurements of the total
conductivity, $\sigma_T$, but more can be said if both $\sigma$
and $\tilde\sigma$ are separately known. We now describe the
implications of the symmetry for one such a measurement.

Ref.~\cite{Kellogg} examines a crossover between two different
plateaux -- one with total filling factor $\nu_T = 2$ and the
other with $\nu_T = 1$ -- and explicitly give enough information
to determine separately all of the complex components of the
resistivity tensor, $\rho$ and $\tilde \rho$ as follows. Figure 9
reproduces the four resistance traces, $A$, $B$, $C$ and $D$,
given in ref.~\cite{Kellogg}, where the sign of trace $C$ is
reversed. Traces $B$ and $C$ directly give the inter-layer
resistivities,\footnote{The conversion from resistance to
resistivity is immediate since the sample used in
ref.~\cite{Kellogg} is square.}
\begin{equation}
    \rho_B=\tilde\rho_{xy} \qquad  \hbox{and} \qquad
    \rho_C=-\tilde\rho_{xx}
\end{equation}
while $A$ and $D$ represent intra-layer resistivities,
\begin{equation}
\rho_D=\rho_{xy} \,.
\end{equation}


\vtop{ \includegraphics{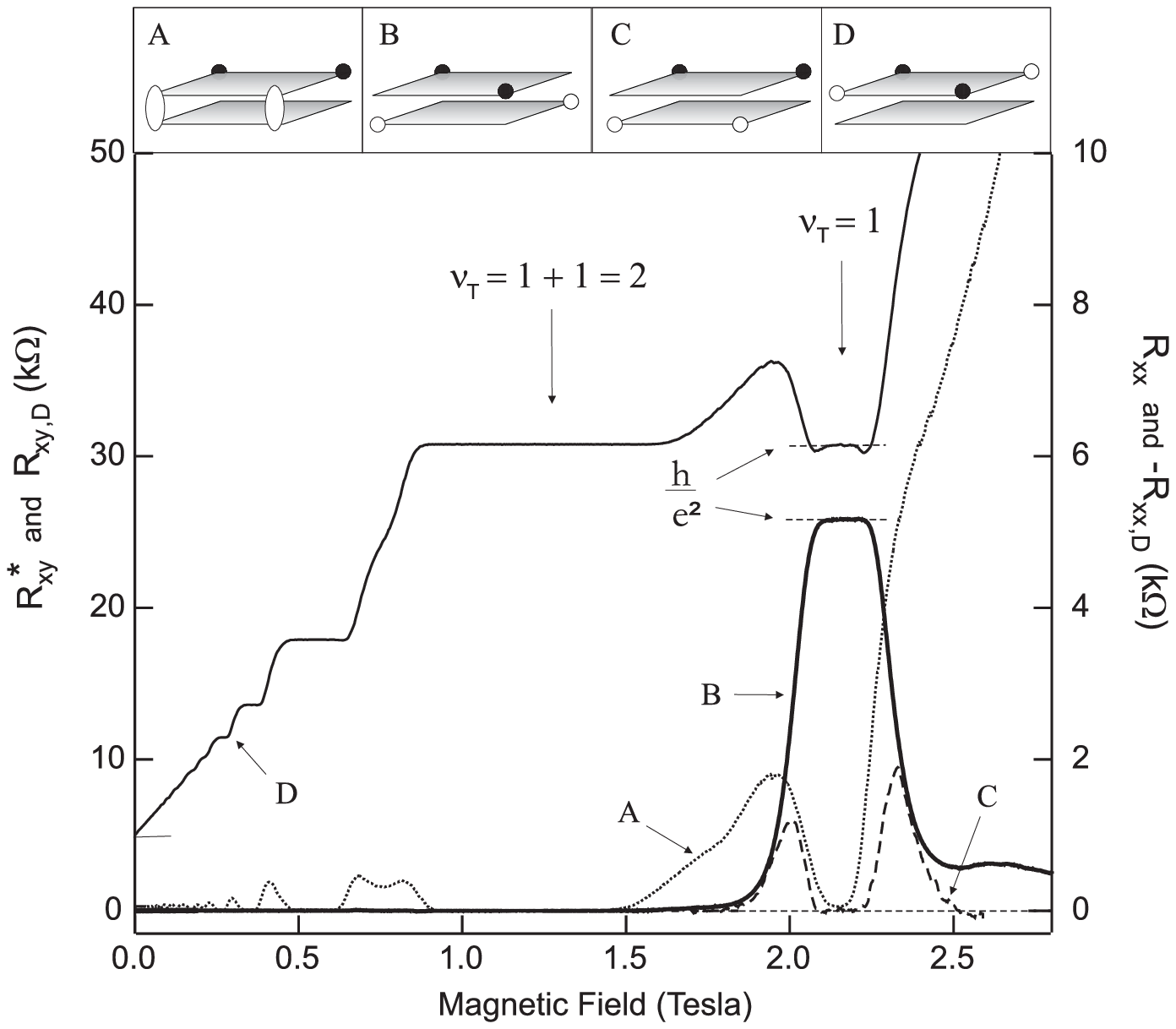} \vskip 8.2cm \centerline{\small
Figure 8: Resistance traces for the bilayer} \vskip 0.005cm
\centerline{\small systems reproduced from Kellogg {\it et.al.}
\cite{Kellogg}.} \vskip 0.005cm \centerline{\small The sign of
trace C is reversed in this plot.}}
\bigskip

The extraction of the intra-layer Ohmic resistivity, $\rho_{xx}$,
from the data is a little less direct. Denoting the $2 \times 2$
complex resistivity tensor by
\begin{equation}
-\Sigma^{-1} = R = \pmatrix{\rho & \tilde\rho\cr \tilde\rho & \rho
\cr}
\end{equation}
then the eigenvalues of $R$,
\begin{equation}
\rho_\pm =\rho\pm\tilde\rho \,,
\end{equation}
are related to the corresponding eigenvalues of $\Sigma$ by
\begin{equation}
\sigma_\pm = \sigma \pm \tilde\sigma = -\frac{1}{\rho_\pm} \,.
\end{equation}
Trace $A$ of Fig.~8 then measures the Ohmic resistance
corresponding to the total conductivity
\begin{equation}
\sigma_T=2(\sigma+\tilde\sigma) = 2\sigma_+ = - \frac{2}{\rho_+} ,
\end{equation}
Consequently $\rho_T = -1/\sigma_T = \rho_+/2$, and so trace $A$
gives $\rho_A=\rho_{+ xx}/2$, or
\begin{equation}
\rho_{xx}=2\rho_A+\rho_C \,.
\end{equation}

Given these resistivities, we see that the two plateaux observed
in \cite{Kellogg} occur for
\begin{equation}
R_1 = \pmatrix{1 & 0\cr 0 & 1 \cr} \label{1010} \qquad \hbox{and}
\qquad R_2 = \pmatrix{1 & 1\cr 1 & 1 \cr} \label{1111} \,,
\end{equation}
which respectively correspond to $(\rho_+,\rho_- )_1 = ( 1,1)$ and
$(\rho_+,\rho_-)_2 = (2,0)$, and so $(\sigma_+,\sigma_-)_1 =
(-1,-1)$ and $(\sigma_+,\sigma_-)_2 = \left( -\frac12,\infty
\right)$. Notice that $R_2$ is not invertible.

According to our proposal, the action of $Sp(4,Z)$ on the bilayer
system should manifest itself as $\Gamma_0(2)$ acting separately
on $\sigma_+$ and $\sigma_-$, with implications $B1-B3$ or
$B1'-B3'$, depending on the category of flows between phases in
which the system lies. For $\sigma_\pm = p_\pm/q_\pm$ prediction
$B1$ means in particular that $q_\pm$ is odd while $B1'$ implies
$p_\pm$ is odd while $q_\pm$ is even. We see from this that
$\sigma_+ = \sigma_- = -1$ corresponds to condition $B1$ while
$\sigma_+ = - \frac12$ corresponds to $B1'$, and so the transition
observed in ref.~\cite{Kellogg} is {\it between} the two different
categories of bilayer flows.

Because the transition is not purely within one category or
another we have no right to expect either of the sets of predictions
$B1-B3$ or $B1'-B3'$ to apply. Interestingly enough prediction
$B2$ does still seem to apply, since both $\sigma_+: -1 \to
-\frac12$ and $\sigma_-: -1 \to \infty$ satisfy the selection rule
$|pq'-p'q| = 1$ if we interpret $\infty = 1/0$.

\section{Conclusions}

We have seen that the emergence of modular symmetries in
two-dimensional semiconductors has striking observational
consequences both for bilayer and spin-degenerate Quantum Hall
systems, just as it did in the better-established case of
monolayer semiconducting heterostructures. These consequences
include predictions for the pattern of allowed Hall plateaux;
predictions for selection rules restricting which plateaux can be
obtained from which by varying magnetic fields and charge
densities at low temperatures; and, for samples with particle-hole symmetry,
semicircle laws for the
trajectories taken through the conductivity plane during these
transitions. Experimental verification of these predictions would
provide support for the picture that such modular symmetries are
to be expected on very general grounds for two-dimensional
systems. We hope these observations will stimulate experimental
tests of these predictions.

\section*{Acknowledgements}
We thank M. Hilke for helpful discussions. Our research has been
assisted by research funds provided by N.S.E.R.C. (Canada), the
Killam Foundation, McGill and McMaster Universities and Enterprise
Ireland (E.I. Basic Research Grant no. SC/1998/739).
B.D. wishes to thank the Perimeter Institute and C.B. thanks
the Dublin Institute for Advanced Studies for their kind hospitality 
at various times while these ideas were being developed.

\end{document}